\documentclass[namedreferences]{SolarPhysics}
\usepackage{spr-sola-addons} 
\usepackage{epsfig}                     
\usepackage{graphicx}                    
\usepackage{color}                       
\usepackage{url}                         
\usepackage{amsfonts}
\usepackage{lineno}

\begin{document}
\begin{article}
\begin{opening}
\title{Effects of Thomson-Scattering Geometry on White-Light
Imaging of an Interplanetary Shock: Synthetic Observations from
Forward Magnetohydrodynamic Modelling}

\author{Ming~\surname{Xiong}$^{1,2}$\sep
        J. A.~\surname{Davies}$^{3}$\sep
        M. M.~\surname{Bisi}$^{1}$\sep
M. J.~\surname{Owens}$^{4}$\sep %
R. A.~\surname{Fallows}$^{5}$\sep %
G. D.~\surname{Dorrian}$^{6}$\sep
       }

\institute{$^{1}$ Aberystwyth University, Aberystwyth, UK.
                  email: \url{mxiong@spaceweather.ac.cn} \\
           $^{2}$ SIGMA Weather Group, State Key Laboratory for Space Weather, Center for Space Science and Applied Research, Chinese Academy of Sciences, Beijing, China
           \\
           $^{3}$ RAL Space, Harwell Oxford, UK \\
           $^{4}$ Reading University, Reading, UK \\
           $^{5}$ ASTRON, Institute for Radio Astronomy, Netherlands
           \\
           $^{6}$ Institute of Astronomy and Astrophysics, National Observatory of Athens, Lofos Nymfon, Greece %
}

\begin{abstract}
Stereoscopic white-light imaging of a large portion of the inner
heliosphere has been used to track interplanetary coronal mass
ejections. At large elongations from the Sun, the white-light
brightness depends on both the local electron density and the
efficiency of the Thomson-scattering process. To quantify the
effects of the Thomson-scattering geometry, we study an
interplanetary shock using forward magnetohydrodynamic simulation
and synthetic white-light imaging. Identifiable as an inclined
streak of enhanced brightness in a time-elongation map, the
travelling shock can be readily imaged by an observer located
within a wide range of longitudes in the ecliptic. Different parts
of the shock front contribute to the imaged brightness pattern
viewed by observers at different longitudes. Moreover, even for an
observer located at a fixed longitude, a different part of the
shock front will contribute to the imaged brightness at any given
time. The observed brightness within each imaging pixel results
from a weighted integral along its corresponding ray-path. It is
possible to infer the longitudinal location of the shock from the
brightness pattern in an optical sky map, based on the east-west
asymmetry in its brightness and degree of polarization. Therefore,
measurement of the interplanetary polarized brightness could
significantly reduce the ambiguity in performing three-dimensional
reconstruction of local electron density from white-light imaging.
\end{abstract}


\end{opening}

\section{Introduction}
Interplanetary space is permeated with the supersonic solar wind
flow from the Sun. The ubiquitous solar wind not only carries
large-amplitude Alfv\'{e}n waves \cite{Belcher1971,Li2007} but
also is a transmission medium for interplanetary coronal mass
ejections (CMEs) \cite{Dryer1994}. CMEs, firstly discovered by
white-light coronagraphs in the 1970s, are large-scale expulsions
of plasma and magnetic field from the solar atmosphere. A typical
CME carries a magnetic flux of $10^{23}$ Maxwells and a plasma
mass of $10^{16}$ grams \cite{Forbes2006}. During interplanetary
propagation, a CME may interact with the background structures in
the ambient solar wind, such as the heliospheric current sheet
(HCS) and corotating interaction regions (CIRs). Moreover, at
solar maximum, multiple CMEs successively launched from the Sun
are likely to interact with each other. The interaction is
generally significant, nonlinear, and irreversible. The resultant
complex travelling structures, sampled later at high temporal
resolution by {\it in situ} instrumentation at 1 AU, are
identified as magnetic cloud boundary layers \cite{Wei2003},
CME-CIR events \cite{DalLago2006}, CME-shock events
\cite{Lepping1997}, CME-CME events \cite{Burlaga1987,Dasso2009},
and so on. Interplanetary CMEs have been widely recognized as the
primary driver of interplanetary disturbances and large
geomagnetic storms ({\it e.g.}, \opencite{Burlaga1987};
\opencite{Gosling1991}; \opencite{Lepping1997};
\opencite{DalLago2006}).

Optical heliospheric imagers (HIs) enable continuous monitoring of
the evolution of such solar eruptions as they propagate through
interplanetary space. Heliospheric imaging fills the observation
gap between near-Sun coronagraph imaging and {\it in situ}
measurements. Interplanetary structures are viewed by means of
Thomson-scattered sunlight. The scattered light retains the same
spectral characteristics as the incident sunlight on the basis of
Thomson-scattering theory \cite{Billings1966,Howard-T-A2009a}. The
effects of multiple scattering are negligible, because the solar
wind is optically thin. The angle between the Sun and a target,
such as a CME, as viewed from an observer, is termed elongation
$\varepsilon$. The brightness difference between the Sun and a
target (to be observed) at a large elongation is many orders of
magnitude. More specifically, large CMEs at $\varepsilon=45^\circ$
have optical intensities that are of order $10^{-14} \, {\rm
B}_\odot$ \cite{DeForest2011}, where ${\rm B}_\odot$, the solar
brightness unit, is the intensity of the Sun as a power per unit
area (of the photosphere) per unit solid angle. To map
interplanetary structures at large elongations, the
Thomson-scattered signals must be separated from many other
sources of diffuse light: background light from the Sun, zodiacal
light, and starlight (c.f., Figure 3 from \opencite{Harrison2005};
Figures 5 from \opencite{Eyles2009}; Figure 1 from
\opencite{Jackson2010}). The Thomson-scattering signals are faint
and transient, whereas the other signals are intense but,
fortunately, stable \cite{Leinert1989}. So by successfully
subtracting the background brightness, an optical sky map taken by
an HI easily reveals interplanetary transients irradiated by
direct sunlight ({\it e.g.}, \opencite{DeForest2011}).
\opencite{DeForest2011} proposed an image processing procedure
consisting of five major steps: stationary background removal,
celestial background removal (including cross-image distortion
measurement), residual F corona removal, moving feature filtration
in the Fourier plane, and conversion back to focal plane
coordinates. Using this procedure, the solar wind at 1 AU can be
imaged with a sensitivity of a few $\times 10^{-17} {\rm
B}_\odot$, compared to a background signal of a few $\times
10^{-13} {\rm B}_\odot$. Instrument specifications for an HI
require careful design that incorporates the stray-light impacts
of the HI bus, HI appendages, and other instruments onboard the
spacecraft \cite{Harrison2005,Eyles2009,Jackson2010}. Historic
milestones in terms of spaceborne white-light imaging were
achieved by the zodiacal-light photometers \cite{Leinert1981}
onboard the {\it Helios} spacecraft, the Large Angle and
Spectrometric COronagraph ({\it LASCO}) \cite{Brueckner1995}
onboard the Solar and Heliospheric Observatory ({\it SOHO}), the
Solar Mass Ejection Imager ({\it SMEI}) \cite{Eyles2003} onboard
the {\it Coriolis} spacecraft, and the {\it HIs}
\cite{Howard2008,Harrison2008,Eyles2009} onboard the twin
Solar-TErrestrial RElations Observatory ({\it STEREO})
\cite{Kaiser2008}. The {\it STEREO} mission is comprised of two
spacecraft, with one leading the Earth in its orbit ({\it STEREO
A}) and the other trailing behind the Earth ({\it STEREO B}). Both
{\it STEREO A} and {\it B} separate from the Earth by $22.5^\circ$
per year. The HI instrument on each {\it STEREO} spacecraft
consists of two cameras, {\it HI}-1 and {\it HI}-2, whose optical
axes lie in the ecliptic. Elongation coverage in the ecliptic is
$4^\circ$ -- $24^\circ$ for {\it HI}-1 and $18.7^\circ$ --
$88.7^\circ$ for {\it HI}-2; The field-of-view (FOV) is $20^\circ
\times 20^\circ$ for {\it HI}-1 and $70^\circ \times 70^\circ$ for
{\it HI}-2; The cadence is 40 minutes for {\it HI}-1 and 2 hours
for {\it HI}-2 \cite{Eyles2009,Harrison2009}. With the launch of
{\it STEREO}, a CME can be imaged from its nascent stage in the
corona all the way out to 1 AU and beyond, including those that
are Earth-directed
\cite{Harrison2008,Davies2009,Liu2010,Lugaz2010,DeForest2011}.
Moreover, images from the {\it STEREO}/{\it HI}-2 have revealed
detailed spatial structures within CMEs, including leading-edge
pileup, interior voids, filamentary structure, and rear cusps
\cite{DeForest2011}. The leading-edge pileup of solar wind
material is observed as a bright arc in optical imaging and
revealed as a shock sheath from {\it in situ} sampling. With the
launch of the {\it STEREO} mission, the interaction between CMEs
in the inner heliosphere can be simultaneously imaged from
multiple vantage points \cite{Liu2012}. HIs are planned as part of
the payload on a number of future missions, including the {\it
Solar Orbiter} and the {\it Solar Probe Plus}.

White-light images from HIs are more difficult to interpret than
those from coronagraphs. CMEs imaged by the {\it STEREO}/{\it HIs}
are often faint and diffuse. By stacking a time series of running
difference brightness along a fixed spatial slice, often
corresponding to the ecliptic, a time-elongation map (commonly
called a J-map) is produced. In such a J-map, the signatures of
faint propagating transients are highlighted, and the transient
can be tracked. However, the interpretation of the leading edge of
the transient track in a J-map can be quite controversial. By
fitting its time-elongation profile extracted from the J-map
created from single-spacecraft observations
\cite{Sheeley2008,Rouillard2008}, the radial velocity and
direction of a CME can be estimated, assuming that the CME moves
radially at a constant velocity. Such a fit can be performed
independently for the two {\it STEREO} spacecraft, if the CME is
observed by both. Moreover, if it is observed by both spacecraft,
the CME's position (and hence its velocity) and its propagation
direction can be inferred as a function of time throughout its
observation by direct triangulation \cite{Liu2010} or a ``tangent
to a sphere" method \cite{Lugaz2010}. The determination of CME
kinematics, in particular the propagation direction, can be
ambiguous. The assumptions that one is always observing the same
part of a CME can lead to large errors in the estimated height of
the CME's leading edge. The structure of a CME affects the
derivation of its kinematic properties, particularly at large
elongations. As a CME propagates to increasingly large
elongations, the location of its leading edge changes. As stated
by \opencite{Howard-T-A2009a}, the interpretation of observations
of the leading edge of a CME can be fraught with difficulties,
especially at large elongations. Scattered sunlight in the inner
heliosphere is not only determined by the local electron density,
but also by Thomson-scattering theory. The Thomson-scattering
geometry is defined by a sphere, on which the Sun and the observer
are located on opposite ends of the diameter. The classical theory
of Thomson-scattering, by \opencite{Billings1966}, was revisited
by \opencite{Howard-T-A2009a}. \opencite{Howard-T-A2009a}
concluded that: (1) The result of the scattering efficiency
contribution is to somewhat de-emphasize the importance of the
Thomson-scattering surface; (2) The Observed intensity is
distributed out to large distances from the Thomson-scattering
surface; (3) Such a spread of observed intensity is more
significant at larger elongations. A CME can, therefore, be
readily imaged at large elongations. Using SMEI observations,
\opencite{Howard2007} demonstrated that limb CMEs could be
detected out to at least 0.5 AU. Furthermore,
\opencite{Tappin2009} showed that CMEs could be phenomenologically
modelled as a modified bubble or shell structure. However,
inferring the three-dimensional location of a CME becomes more
difficult with increasing elongation. In optical imaging, the
observed intensity is traditionally assumed to come from a
so-called plane-of-sky (POS). The POS is defined as the plane
containing the Sun and being perpendicular to the line-of-sight
(LOS) from an observer. While such a simple assumption is
reasonable for coronagraph imaging, it is invalid for heliospheric
imaging \cite{Howard-T-A2009a,Jackson2010}. An inappropriate
application of the POS assumption to heliospheric imaging would
result in a significant underestimation of CME mass. A CME is
inherently three-dimensional in nature, and imaging observations
only provide two-dimensional information as a result of LOS
integration through the three-dimensional structure. Effects of
the projection and the Thomson-scattering are likely to affect
brightness tracks in time-elongation maps. Uncertainties resulting
from such effects can be assessed and disentangled quantitatively
using a magnetohydrodynamic (MHD) modelling as a digital
laboratory. For instance, CMEs on 24 -- 25 January 2007 were
modelled, and synthetic optical images compared in detail with
observations from the {\it STEREO}/{\it HIs}, by
\opencite{Lugaz2009}. Moreover, by forward modelling of a
travelling shock, \opencite{Xiong2011} demonstrated that
remote-sensing signatures in coordinated white light and
interplanetary scintillation (IPS) observations can be
theoretically simulated. Numerical models are useful in
interpreting the brightness patterns of CMEs tracked in optical
sky maps, and hence aiding our understanding of heliospheric
processes such as the propagation, evolution, and possible
interactions of CMEs.

In this paper, we simulate a shock propagating through the inner
heliosphere using an MHD model. Subsequently, we synthesize and
investigate the corresponding white light images that would be
observed from 1 AU. We present the numerical MHD model and the
optical Thomson-scattering formulation in Section
\ref{Sec:Method}. In Section \ref{Sec:LOS}, we describe tracking
of the shock front from different perspectives, and present
profiles of the scattered sunlight intensity along various LOSs.
Subsequently, we analyze the brightness patterns observed in
synthetic time-elongation maps and longitude-elongation maps
(Section \ref{Sec:Pattern}). Finally, we discuss remote sensing of
much more complex interplanetary phenomena in both white light and
radio in Section \ref{Sec:Summary}.

\section{Method}\label{Sec:Method}
A numerical model of the heliosphere \cite{Xiong2006a} is used to
associate interplanetary dynamics with corresponding observed
signatures. Such a self-consistent link is summarized in the flow
chart for synthetic remote-sensing in white light and radio shown
in Figure 8 from \opencite{Xiong2011}. Here, we investigate only
white-light imaging for a travelling fast shock. The procedure
consists of two steps, described in Sections \ref{Sec-sub:MHD} and
\ref{Sec-sub:TS}.

\subsection{Numerical MHD Model}\label{Sec-sub:MHD}
Large-scale dynamics of the inner heliosphere can be physically
described by MHD processes. Using a sophisticated numerical MHD
model \cite{Xiong2006a}, a blast shock wave travelling through
interplanetary space is numerically simulated in this paper. We
describe our model in Table \ref{Tab:model}, establish the
background solar wind in Table \ref{Tab:BG}, and prescribe the
shock injection in Table \ref{Tab:Shock}. The simulated electron
density is used to generate synthetic white-light images.

\subsection{Thomson-Scattering Calculations} \label{Sec-sub:TS}
\begin{figure}
\noindent
\includegraphics[width=0.95\textwidth]{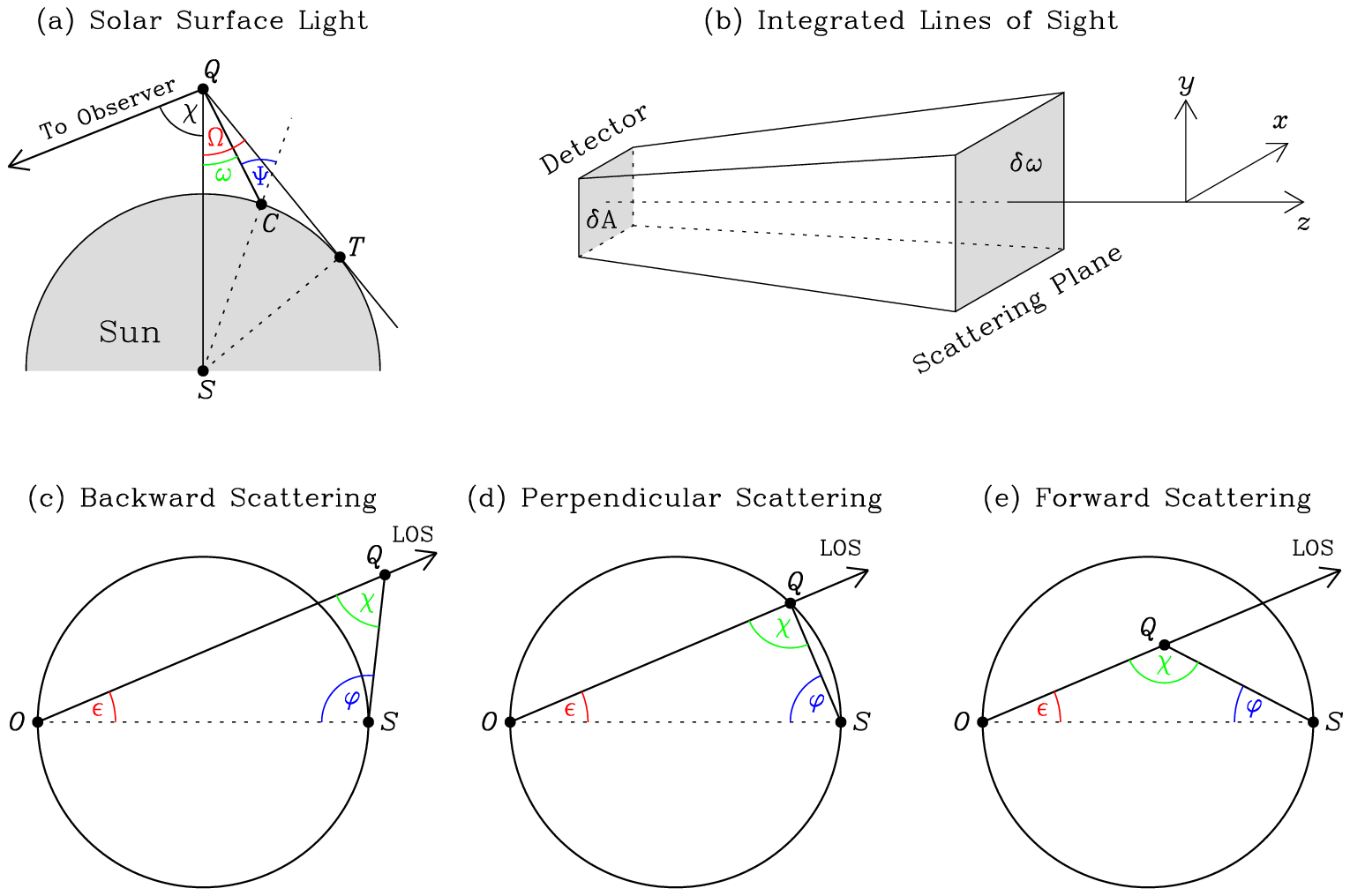}
\caption{} \label{Cartoon}
\end{figure}

White-light imaging is performed on the basis of the
well-established Thomson-scattering theory. Interplanetary free
electrons scatter white-light photons from the photosphere, so the
inner heliosphere is visible as a brightness sky map. In this
paper, we adopt the formulation of Thomson-scattering theory in
interplanetary imaging given by \opencite{Howard-T-A2009a}. This
is demonstrated schematically in Figure \ref{Cartoon}. At a
scattering point $Q$, the solar surface is only visible within a
cone where $\angle SQT =\Omega$ and $\angle STQ =90^\circ$. The
Sun looks like a limb-darkened luminous disk. All straight rays of
direct sunlight intercepted at point $Q$ are scattered into a full
solid angle of $4 \, \pi$, with part of the sunlight being
scattered toward an observer at point $O$. The observer $O$,
scattering point $Q$, and Sun center $S$ all lie in an emergent
plane. At the observation site $O$, received sunlight is direct
and unpolarized along the LOS $SO$, and scattered and polarized
along the LOS $QO$. The luminous magnitude of the Sun (LOS $SO$)
is dramatically higher than that of interplanetary space (LOS
$QO$) \cite{Harrison2005,Eyles2009,Jackson2010,DeForest2011}. A
series of baffles onboard the {\it STEREO}/{\it HIs}
\cite{Eyles2009} attenuates the intense solar radiation to such a
degree that the much fainter interplanetary brightness is
revealed. Scattered by a single electron at point $Q$, the optical
intensity $G$, measured at point $O$ as a power per unit solid
angle, can be expressed in units of ${\rm B}_\odot$. The intensity
$G$ is polarized along the LOS $QO$, and can be expressed in terms
of two orthogonal components $G_{\rm R}$ and $G_{\rm T}$. The
radial component $G_{\rm R}$ is in the previously defined emergent
plane (the $OQS$--plane) and is perpendicular to the LOS $QO$. The
tangent component $G_{\rm T}$ is perpendicular to the
$OQS$--plane. Mathematical expressions for the luminous intensity
$G$ and its constituent components ($G_{\rm R}$ and $G_{\rm T}$)
are formulated in terms of van de Hulst coefficients $A$, $B$,
$C$, and $D$ \cite{Howard-T-A2009a}:
\begin{eqnarray}
  &&G_{\rm P}= G_{\rm T} - G_{\rm R} \nonumber\\
  &&G= G_{\rm T} + G_{\rm R} = 2 \,G_{\rm T} - G_{\rm P} \nonumber\\
  &&G_{\rm T}= \frac{\pi \, \sigma_{\rm e}}{2\,z^2} \,[(1-u) C + uD] \label{Equ:I-T} \\
  &&G_{\rm P}= \frac{\pi \, \sigma_{\rm e}}{2\,z^2} \, \sin^2\chi \, [(1-u) A +
  uB] \nonumber
\end{eqnarray}
Here $G_{\rm P}$ is an auxiliary parameter. $\sigma_{\rm e}$,
defined as the differential cross section for perpendicular
scattering, is a constant of $7.95 \times 10^{-30}$ m$^2$
Sr$^{-1}$ \cite{Howard-T-A2009a}. $u$, defined as a limb-darkened
coefficient, is a function of wavelength. A value for $u$ of 0.63,
for a wavelength of 5500 \AA $\,$ is adopted in this paper. $z$ is
the distance from the scattering point to the observer. The
coefficients $A$, $B$, $C$, and $D$ \cite{Billings1966} are given
as:
\begin{eqnarray}\label{Equ:ABCD}
  &&A=\cos \Omega \, \sin^2 \Omega \nonumber \\
  &&B=-\frac{1}{8} \, \left [1-3\, \sin^2 \Omega - \frac{\cos^2 \Omega}{\sin
  \Omega} (1 + 3\, \sin^2 \Omega) \ln \left (\frac{1+\sin \Omega}{\cos
  \Omega} \right ) \right ] \nonumber \\
  &&C=\frac{4}{3} - \cos \Omega - \frac{\cos^3 \Omega}{3} \\
  &&D=\frac{1}{8} \, \left [5 +  \sin^2 \Omega - \frac{\cos^2 \Omega}{\sin \Omega} (5-\sin^2 \Omega) \ln \left (\frac{1+\sin \Omega}{\cos
  \Omega}\right ) \right ]  \nonumber
\end{eqnarray}
As the scattering point $Q$ increases in heliocentric distance
$r$, $\angle SQT =\Omega$ becomes smaller and the Sun tends toward
a point source. Accordingly, the coefficients $A$, $B$, $C$, and
$D$ fall off as a function of $r^{-2}$, and the differences given
by $|A-C|$ and $|B-D|$ reduce. As a result, the incident sunlight
becomes fainter and more collimated (c.f., Figure 4 from
\opencite{Howard-T-A2009a}).

An imaging detector collects photons that fall within its FOV, and
the brightness within a pixel will comprise contributions from all
scattering sites along its LOS. \opencite{Howard-T-A2009a} pointed
out, however, that ``{\it When interpreting real observations, it
is crucial to note that although the scattering is presented as a
line-of-sight integral this is not strictly true. It is in fact an
integral though the cone of the instrument's point spread
function.}". The configuration for the LOS integral calculation is
demonstrated schematically in Figure \ref{Cartoon}b. The
scattering plane is at a distance $z$ from the detector. The
detector itself has an area of $\delta A$, and its FOV at distance
$z$ has an area given by $\delta \omega = dx \, dy$. The solid
angle subtended by the detector at the scattering area $\delta
\omega$ at $z$ is given by $d \omega= \delta \omega / z^2 =dx \,
dy / z^2$. The brightness contribution from distance $z$ to $z+dz$
along the LOS, where the electron number density is given by $n$,
can be expressed as follows:
\begin{equation}
 \begin{array}{l}
 \left( \begin{array}{c}
   i\\
   i_{\rm P}\\
   i_{\rm T}\\
   i_{\rm R}\\
 \end{array}\right)dz =
 \displaystyle \iint_{\delta \omega} n \, {\rm B}_\odot
\left( \begin{array}{c}
   G\\
   G_{\rm P}\\
   G_{\rm T}\\
   G_{\rm R}\\
 \end{array}\right) dx \, dy\, dz \, \delta A =
   n \, z^2
\left( \begin{array}{c}
   G\\
   G_{\rm P}\\
   G_{\rm T}\\
   G_{\rm R}\\
 \end{array}\right) {\rm B}_\odot \, d\omega \, \delta A\, dz
 \end{array}.\label{Equ:LOS-orig}
\end{equation}
If we let \begin{equation} {\rm B}_\odot \, d\omega \, \delta A =
1.
\end{equation}
Equation (\ref{Equ:LOS-orig}) can be further simplified to
\begin{equation}
 \begin{array}{l}
 \left( \begin{array}{c}
   i\\
   i_{\rm P}\\
   i_{\rm T}\\
   i_{\rm R}\\
 \end{array}\right) =
   n \, z^2
\left( \begin{array}{c}
   G\\
   G_{\rm P}\\
   G_{\rm T}\\
   G_{\rm R}\\
 \end{array}\right)
 \end{array}.\label{Equ:LOS-dot}
\end{equation}
Within each pixel of the detector, the optical brightness that is
recorded is the LOS integral of Equation (\ref{Equ:LOS-dot}):
\begin{equation}
 \begin{array}{l}
 \left( \begin{array}{c}
   I\\
   I_{\rm P}\\
   I_{\rm T}\\
   I_{\rm R}\\
 \end{array}\right) =
\displaystyle \int^\infty_0 \left( \begin{array}{c}
   i\\
   i_{\rm P}\\
   i_{\rm T}\\
   i_{\rm R}\\
 \end{array}\right) dz=
  \displaystyle \int^\infty_0 n \, z^2
\left( \begin{array}{c}
   G\\
   G_{\rm P}\\
   G_{\rm T}\\
   G_{\rm R}\\
 \end{array}\right) dz
 \end{array}.\label{Equ:LOS}
\end{equation}
Hence it can be seen that the electron density $n$ and
Thomson-scattering geometry factors ($z^2 G$, $z^2 G_{\rm R}$,
$z^2 G_{\rm T}$) jointly determine the observed brightness. In
terms of the electric field oscillations $I_{\rm R}$ and $I_{\rm
T}$, polarization $p$ is defined as:
\begin{eqnarray}\label{Equ:Polarization}
 && p=\frac{I_{\rm T} - I_{\rm R}}{I_{\rm T} + I_{\rm R}} = \frac{I_{\rm
 P}}{I}.
\end{eqnarray}

Sunlight is scattered backward for $\chi <90^\circ$ (Figure
\ref{Cartoon}c), perpendicular for $\chi =90^\circ$ (Figure
\ref{Cartoon}d), and forward for $\chi
>90^\circ$ (Figure \ref{Cartoon}e).
Perpendicular scattering toward a fixed observer $O$ comes from
the so-called Thomson-scattering sphere, which is centered between
the Sun and the observer (Figure \ref{Cartoon}d). On the
Thomson-scattering surface itself, Thomson-scattering is actually
minimized due to perpendicular scattering. However, both incident
sunlight intensity and local electron density are maximized on the
Thomson-scattering surface, that being where any LOS is closest to
the Sun. These three competing effects determine the scattered
sunlight intensity. Although maximized on the Thomson-scattering
surface itself, the scattered intensity is distributed with
distance away from the surface \cite{Howard-T-A2009a}. Owing to
the weakest scattering efficiency being on the Thomson-scattering
sphere, the importance of the Thomson-scattering sphere is
de-emphasized, particulary at large elongations. On the
Thomson-scattering sphere, the radial component $I_{\rm R}$ of
scattering is the smallest and tangent component $I_{\rm T}$ is
the largest. Thus, the polarization of scattered sunlight is
largest on the Thomson-scattering sphere, according to Equation
(\ref{Equ:Polarization}). If it can be measured at large
elongations, the polarization of scattered sunlight can provide an
important clue to identifying corresponding major scattering
location.

\section{Line-of-Sight Profiles of Scattered Sunlight}\label{Sec:LOS}
\begin{figure}
\noindent
\includegraphics[width=0.95\textwidth]{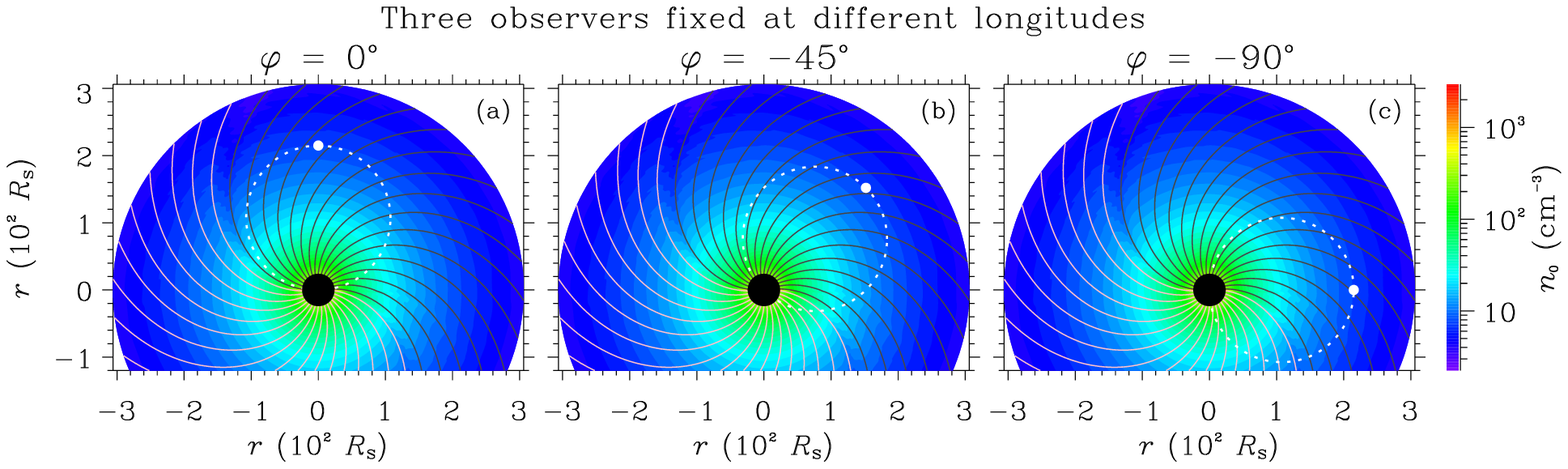}
\caption{} \label{init}
\end{figure}

\begin{figure}
\noindent
\includegraphics[width=0.95\textwidth]{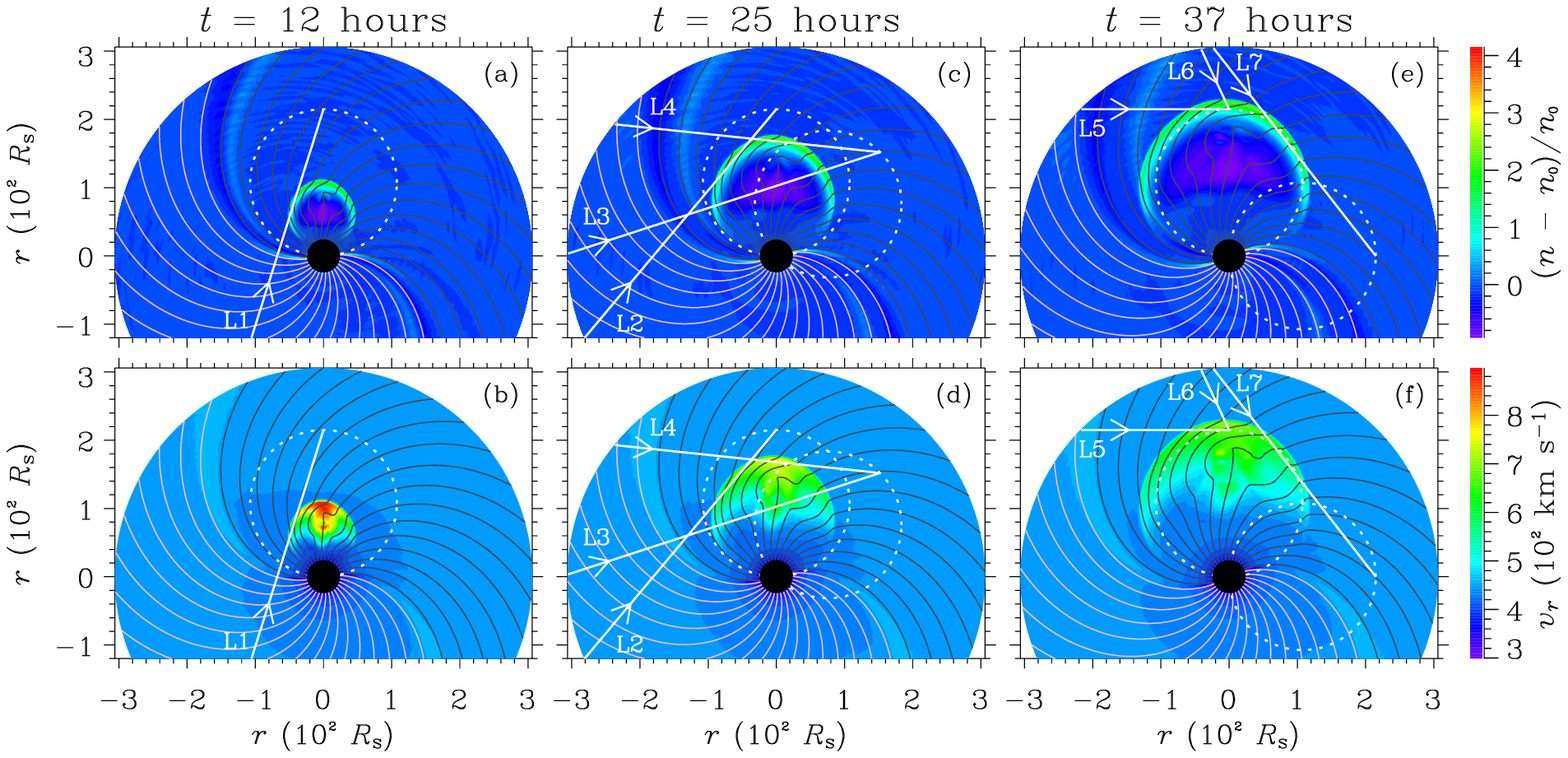}
\caption{} \label{Contour}
\end{figure}

The position of an interplanetary shock can be identified from a
white-light imaging map as a locally enhanced brightness pattern.
Sunlight scattered from a travelling shock is more intense than
that scattered by the ambient solar wind, as the electrons are
significantly compressed just downstream of the shock front. For
the purposes of this paper, we consider an interplanetary shock
viewed simultaneously by three observers that are located at fixed
longitudes of $\varphi=0^\circ$, $-45^\circ$, $-90^\circ$. The
observers and their corresponding Thomson-scattering spheres are
depicted in Figure \ref{init}. The observed brightness pattern
looks different when viewed from the three different sites, and
its leading edge corresponds to different sections of the shock.
As shown in Figure \ref{Contour}, the nose, left flank, and right
flank of the shock are penetrated by LOSs L1, L4, and L7,
respectively. Ray-paths L1, L4, and L7 are directed toward
different observers, at longitudes of $\varphi=0^\circ$,
$-45^\circ$, $-90^\circ$, respectively. Each of the ray-paths L1,
L4, and L7 corresponds to the largest elongation at which enhanced
brightness is observed in a full sky map. Even for a fixed
observer, the leading edge of the brightness pattern actually
corresponds to a different section of the shock at a given time.
This can be demonstrated by considering an observer-directed shock
travelling along the longitude $\varphi=0^\circ$ (Figure
\ref{Contour}). The travelling shock front is intersected by LOS
L1 at 12 hours, L2 at 25 hours, and L6 at 37 hours, respectively.
The shock flank contributes to the brightness pattern observed at
12 hours, and the shock nose at 37 hours. Such a transition from
the flank to the nose is smooth in time. With respect to the
observer, plasma packets along a single LOS generally have
different spatial speeds (where a spatial speed is expressed in an
observer-centered polar coordinate system). Different spatial
speeds generally correspond to different observer-centered angular
speeds. So, plasma packets temporarily aligned along one LOS will
spread over adjoining LOSs later (c.f., Figure 6 from
\opencite{Xiong2011}). It should be remembered that when one
visualizes optical signatures of an interplanetary CME in a
time-elongation map, he/she is not always observing the same part
of the structure. Interpretations of brightness patterns at large
elongations are usually ambiguous.

\begin{figure}
\noindent
\includegraphics[width=0.96\textwidth]{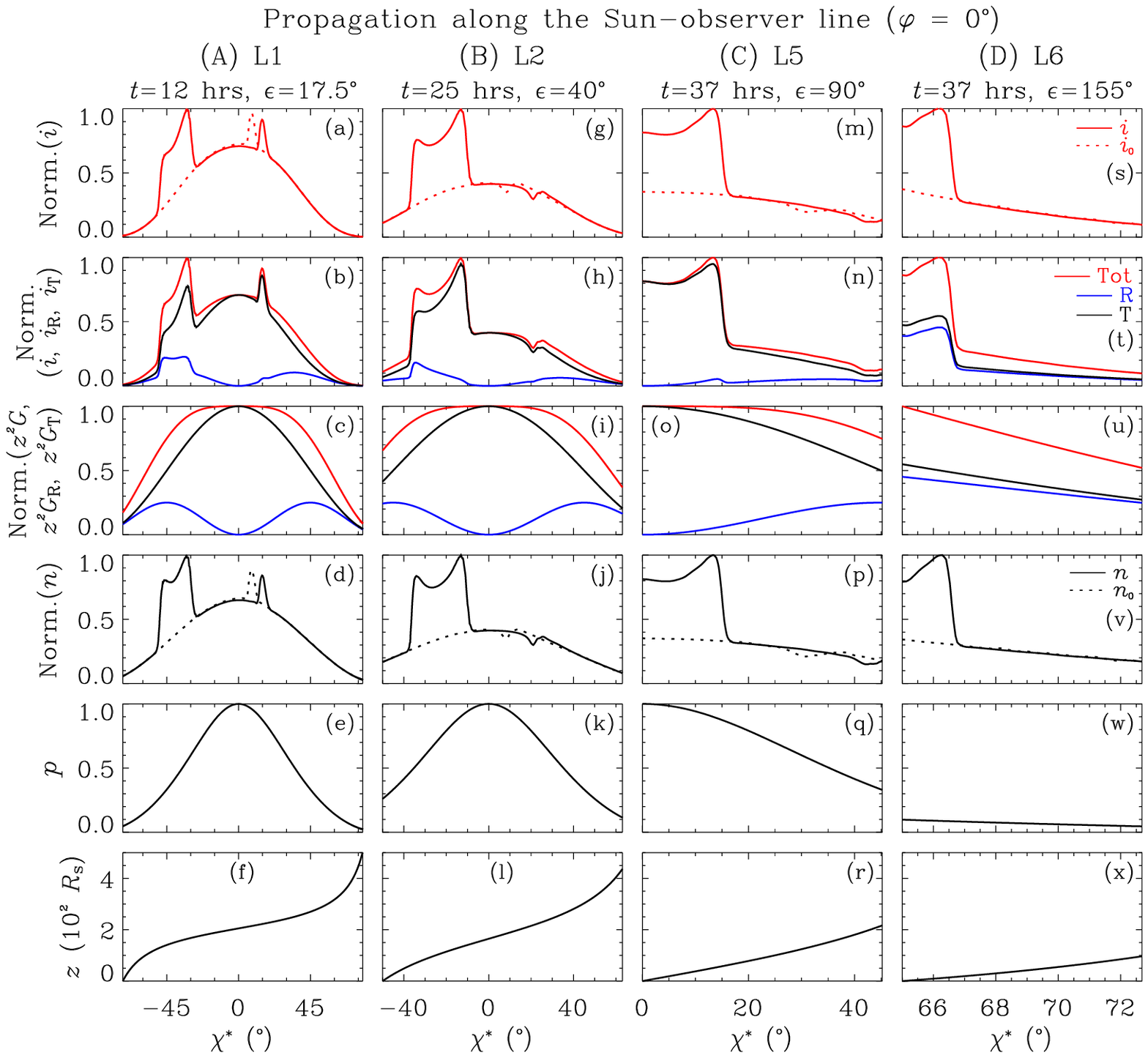}
\caption{} \label{LOS-a}
\end{figure}

\begin{figure}
\noindent
\includegraphics[width=0.96\textwidth]{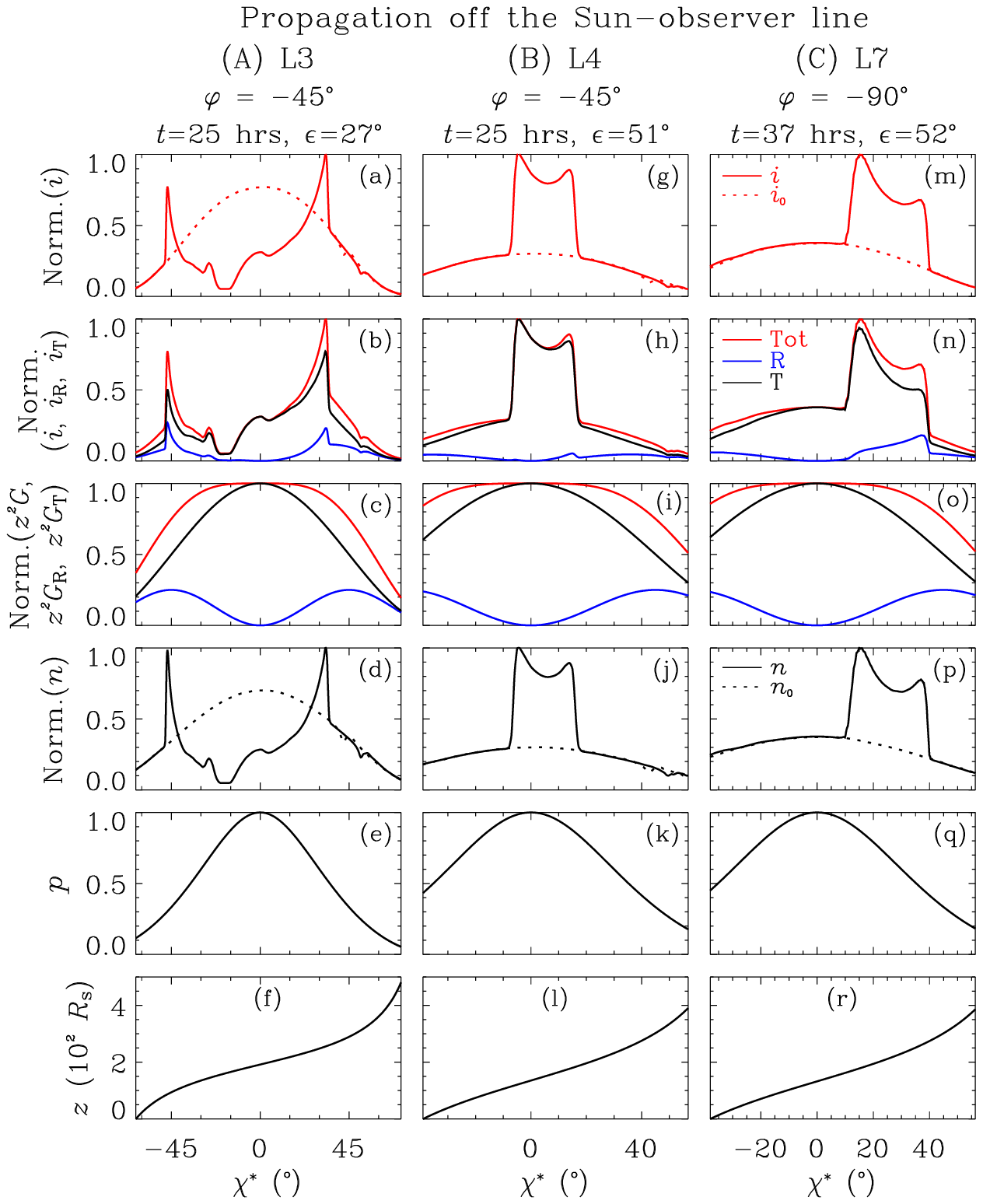}
\caption{} \label{LOS-b}
\end{figure}

Sunlight scattered along an LOS is collected in the corresponding
pixel and recorded in terms of optical brightness. The brightness
in the pixel results from an integral along the LOS, as expressed
in Equation (\ref{Equ:LOS}). Thus, the LOS distributions of
localized intensities ($i$, $i_{\rm R}$, $i_{\rm T}$) and
Thomson-scattering geometry factors ($z^2 \, G$, $z^2 \, G_{\rm
R}$, $z^2 \, G_{\rm T}$) will determine which part of the LOS
contributes most significantly to the total intensities observed
($I$, $I_{\rm R}$, $I_{\rm T}$). Such analysis of an
interplanetary shock is performed for LOSs L1--L7 in Figure
\ref{Contour}, and the salient parameters are presented as a
function of modified scattering angle $\chi^*=90^\circ- \chi$ in
Figures \ref{LOS-a} and \ref{LOS-b}. Note that Figures \ref{LOS-a}
and \ref{LOS-b} present normalized values of the intensities,
Thomson-scattering factors, and electron density. The
normalization factors are given in Table \ref{Tab:LOS}. The
Thomson-scattering factors depend on the relative geometry between
the Sun, a scattering site, and the receiving observer. More
specifically, the elongation $\varepsilon$ and scattering angle
$\chi$ determine the Thomson-scattering factors. The dependence of
Thomson-scattering factors on $\varepsilon$ and $\chi$ is explored
by \opencite{Howard-T-A2009a}. For a fixed $\varepsilon$, the
distribution of the Thomson-scattering factors is symmetric around
$\chi^*=0^\circ$ (Figures \ref{LOS-a}c, \ref{LOS-a}i,
\ref{LOS-b}c, \ref{LOS-b}i, and \ref{LOS-b}o). $\chi^*=0^\circ$
corresponds to perpendicular scattering, {\it i.e.,}
$\chi=90^\circ$, and the scattering site lies on the
Thomson-scattering sphere itself (Figure \ref{Cartoon}d). For
$90^\circ < |\varepsilon| \le 180^\circ$, an observer views the
hemisphere opposite the Sun, and all optical paths toward the
observer are backward with $\chi^* > 0^\circ$. One such example is
LOS L6 in Figure \ref{LOS-a}. The profiles of the
Thomson-scattering factors are convex around $\chi^* = 0^\circ$
for $z^2 \, G_{\rm T}$ and concave around $\chi^* = 0^\circ$ for
$z^2 \, G_{\rm R}$ (Figures \ref{LOS-a}c, \ref{LOS-a}i,
\ref{LOS-b}c, \ref{LOS-b}i, and \ref{LOS-b}o). At large
elongations, $z^2 \, G_{\rm R}$ is negligible at $\chi^* =
0^\circ$. Moreover, at $\chi^* = 0^\circ$, the scattered sunlight
is nearly linearly polarized, and its polarization $p$ is almost
1. As elongation $\varepsilon$ increases, the profile of $z^2 \, G
\,-\chi^*$ flattens. Along the nearly anti-sunward LOS L6, where
$\varepsilon = 155^\circ$, the scattered sunlight is almost
unpolarized with $G_{\rm R} \approx G_{\rm T}$ and $p \approx 0$
(Figures \ref{LOS-a}u and \ref{LOS-a}w). The Thomson-scattering
factors remain fairly constant within that region delimited by
$|\chi^*| \le 30^\circ$. In Figure \ref{Contour}, an incident
shock compresses plasma at its front, leaving a low density void
behind. The distributions along LOS L3, which cuts across the
density void and the shock flanks (Figures \ref{Contour}c and
\ref{Contour}d), are presented in Figures \ref{LOS-b}a-f. Figure
\ref{LOS-b}d reveals density spikes within $-49^\circ \le \chi^*
\le -40^\circ$ and $25^\circ \le \chi^* \le 35^\circ$, and a
density void within $-40^\circ < \chi^* < 25^\circ$. The locally
scattered intensity is jointly determined by the electron density
and the Thomson-scattering factors, with $i=n\, z^2 \,G$, $i_{\rm
R}=n\, z^2 \,G_{\rm R}$, and $i_{\rm T}=n\, z^2 \, G_{\rm T}$. The
LOS profile of scattered intensity $i$ (Figure \ref{LOS-b}a) is
similar to that of electron density $n$ (Figure \ref{LOS-b}d);
this is not the case for the background signals. As it is $I$, the
integral of $i$ along the LOS that is recorded by a corresponding
pixel, the two density spikes in Figure \ref{LOS-b}d would be
essentially undetectable, based on measurements along LOS L3.
Other LOSs in Figure \ref{Contour} are, however, tangent to the
shock front. For an observer situated at a longitude
$\varphi=0^\circ$, the travelling shock can be continuously
tracked as a moving brightness pattern. At 12 hours, when the
leading edge of the pattern is observed at an elongation of
$\varepsilon=17.5^\circ$, the shock front is bounded between
$-49^\circ \le \chi^* \le -31^\circ$ along this LOS, L1 (Figure
\ref{LOS-a}d). Simultaneously, a spike-like density enhancement
exists at $\chi^*=15^\circ$; This spike, which is the signature of
the heliospheric plasma sheet (HPS), has shifted from its initial
location of $\chi^*=8^\circ$. The HPS, initially located at
$\varphi=90^\circ$ and with a longitudinal width of only
$3^\circ$, co-rotates with the Sun. The density of HPS reduces
with increasing heliocentric distances, so its optical brightness
is negligible at large elongations. Thus the travelling shock
contributes most significantly to the observed transient
brightness enhancement. As time elapses, the observer-directed
shock is viewed at increasingly larger elongations, and moves
nearer to the corresponding Thomson-scattering sphere.
Accordingly, region in a $n-\chi^*$ plot that corresponds to the
shock compression gradually shifts toward $\chi^*=0^\circ$
(Figures \ref{LOS-a}d, \ref{LOS-a}j, and \ref{LOS-a}p). Once the
observer-directed shock front has passed over the observer at 1 AU
(Figures \ref{Contour}e and \ref{Contour}f), it scatters sunlight
back toward the observer. For such backward scattering, the
Thomson-scattering factors decrease monotonically with an increase
in LOS depth, $z$ (Figures \ref{LOS-a}o and \ref{LOS-a}u). For an
observer viewing along elongations $\varepsilon \ge 90^\circ$,
electrons in the vicinity of the observer contribute to the
majority of observed total brightness, $I$. Hence optical imaging
at elongations $\varepsilon \ge 90^\circ$ can provide measurements
of electron number density in a similar manner to {\it in situ}
observations made by a spaceborne electron detector. For an
observer situated at $\varphi \ne 0^\circ$, the shock is
propagating off the Sun-observer line. However, the shock front is
very likely to cross the observer's corresponding
Thomson-scattering surface. In this case, the most favorable
configuration for white-light imaging is when the nose of the
shock is on the Thomson-scattering surface, as can be seen from
LOS L4 in Figures \ref{Contour}c and \ref{LOS-b}g-l. The wide
flank of the shock wave is readily imaged, even if the shock is
propagating perpendicularly to the Sun-observer line. Imaged from
$\varphi = - 90^\circ$ (Figures \ref{Contour}e and
\ref{Contour}f), the nose and left flank of the shock are too far
away to be detected whereas the right flank of the shock can be
detected along LOS L7 (Figures \ref{LOS-b}m-r). At large
elongations, both the Thomson-scattering sphere and the CME volume
must, therefore, be considered when interpreting optical imaging
of the inner heliosphere.

\section{Optical Brightness Patterns}\label{Sec:Pattern}
Moving brightness patterns in optical sky maps provide
observational evidence of interplanetary transient disturbances
such as travelling CMEs. CMEs often travel faster than the ambient
solar wind, resulting in compression. Thus CMEs are often
associated with dense plasma, such that sunlight scattered within
the CME volume is intense. The wide FOVs of the new generation of
HI instruments encompass a wide range of elongations. Transient
brightness dramatically reduces with an increasing elongation
$\varepsilon$. However, the observed brightness over the entire
FOV can be rescaled to be within the same order of magnitude,
using the normalization factors shown in Figure \ref{calibration}.
In the generation of Figures \ref{calibration}a-d, the electron
number density $n$ is assumed to vary with heliocentric distance
$r$ according to $n \propto r^{-2}$. Intensity profiles ($i$,
$i_{\rm R}$, $i_{\rm T}$) are then calculated using the same
method as for Figures \ref{LOS-a} and \ref{LOS-b}. Assuming that
$n \propto r^{-2}$, and using the known values of the
Thomson-scattering geometry factors ($z^2 G$, $z^2 G_{\rm R}$,
$z^2 G_{\rm T}$), integrated intensities ($I^*$, $I_{\rm R}^*$,
$I_{\rm T}^*$) within $0^\circ \le \varepsilon \le 180^\circ$ are
derived; These are presented in Figures \ref{calibration}e-g.
$I^*$, $I_{\rm R}^*$, and $I_{\rm T}^*$, which are linearly scaled
to be 1 at $\varepsilon=6.7^\circ$, are used to normalize the
intensity $I$, $I_{\rm R}$, and $I_{\rm T}$ shown in Figures
\ref{Longitude} and \ref{Time}. CMEs cause significant derivations
in heliospheric electron density $n$ away from the equilibrium
situation dictated by $n \propto r^{-2}$. These normalized
intensities, $I/I^*$, $I_{\rm R}/I_{\rm R}^*$, and $I_{\rm
T}/I_{\rm T}^*$, can be used to identify transient features in
brightness sky maps. In this paper, we present such normalized
intensities, for the case of a travelling shock (Figure
\ref{Contour}), as longitude-elongation ($\varphi-\varepsilon$)
maps (Figure \ref{Longitude}) and time-elongation
($t-\varepsilon$) maps (Figure \ref{Time}). In Figures
\ref{Longitude}a-i and \ref{Time}a-i, regions with $I/I^* \ge 3.68
\times 10^{-15}$, $I_{\rm R}/I_{\rm R}^* \ge 0.68 \times
10^{-15}$, and $I_{\rm T}/I_{\rm T}^* \ge 2.93 \times 10^{-15}$
are delimited by dotted lines, and are identified as corresponding
to the shock front. The mapped brightness pattern significantly
depends on the longitudinal position of the observer (Figure
\ref{Longitude}). An observer at a location within the range
$150^\circ \le |\varphi| \le 180^\circ$ is nearly opposite to the
direction of shock propagation, so optical signals of the shock
are too faint to be detected. An observer at $\varphi = 0^\circ$
cannot detect the shock until $t=12$ hours. The patterns of
brightness within the longitude-elongation maps are symmetrical
with respect to elongation for $\varphi=0^\circ$ and unsymmetrical
elsewhere (Figure \ref{Longitude}). The propagating shock, as
observed from a fixed longitude, is manifested as an inclined
streak in a time-elongation map (Figure \ref{Time}). The slope of
this feature is steepest for $\varphi=0^\circ$ (Figure
\ref{Time}), as its arrival time at $\varepsilon=100^\circ$ is 35
hours at $\varphi=0^\circ$ compared with 50 hours at
$\varphi=-45^\circ$ and 90 hours at $\varphi=-90^\circ$. The most
intense optical signatures of the shock are those observed from
$\varphi=\pm \, 60^\circ$ at a time of 12 hours (Figure
\ref{Longitude}a) and from $\varphi=\pm \, 30^\circ$ at 25 hours
(Figure \ref{Longitude}b). From these vantage points, the
brightness patterns corresponding to the shock cover the widest
elongation extent and exhibit the largest relative brightness
enhancement due to the close proximity of the shock nose to the
Thomson-scattering surface at these times (see, for example, L4 in
Figure \ref{Contour}c). When the shock front crosses 1 AU, the
optical brightness is enhanced over the entire range of
elongations $90^\circ \le |\varepsilon| \le 180^\circ$,
simultaneously. Such a crossing occurs at around $t=37$ hours for
an observer at $\varphi=0^\circ$ (Figures \ref{Longitude}c and
\ref{Time}a), and near $t=53$ hours for an observer at
$\varphi=-45^\circ$ (Figure \ref{Time}b). The brightness
enhancement, which results from backward scattering of sunlight,
lasts for between 15 and 20 hours (Figures \ref{Time}a-b). These
figures demonstrate how easily an interplanetary shock, with its
wide front, can be optically imaged.

Polarization measurements of scattered sunlight are very useful in
locating the three-dimensional position of a volume of dense
plasma. Viewed along any elongation, the polarized intensities
($I$, $I_{\rm R}$, $I_{\rm T}$) are an LOS integral of the local
electron density, weighted by Thomson-scattering geometry factors.
The Thomson-scattering factors are different for the different
polarized components, {\it i.e.}, $z^2 \, G_{\rm R}$ for a radial
component $i_{\rm R}$ and $z^2 \, G_{\rm T}$ for a tangent
component $i_{\rm T}$ in Equation (\ref{Equ:LOS}). Near the
Thomson-scattering surface (where $\chi^*=0^\circ$), $z^2 \,
G_{\rm T}$ is much larger than $z^2 \, G_{\rm R}$ (Figures
\ref{LOS-a}c, \ref{LOS-a}i, \ref{LOS-a}o, \ref{LOS-b}c,
\ref{LOS-b}i, and \ref{LOS-b}o), and the locally scattered
sunlight is nearly linearly polarized. When the heliospheric
electron density $n$ is in equilibrium, the distribution of $n$
depends only on heliocentric distance $r$, according to the
expression $n \propto r^{-2}$, and the integrated scattered
sunlight along elongation $\varepsilon=70^\circ$ has the largest
polarization, with a value of almost 0.8 (Figure
\ref{calibration}h). An observer viewing sunlight scattered by a
dense parcel of plasma will detect an increase in $I$, an increase
in $I_{\rm T}$, a decrease in $I_{\rm R}$, and an increase in
polarization $p$ as the parcel of plasma approaches the
Thomson-scattering surface. Polarization measurements for the
shock studied in this paper are presented as longitude-elongation
maps in Figure \ref{Longitude} and as time-elongation maps in
Figure \ref{Time}. LOSs L2, L4, L5, and L7 (in Figure
\ref{Contour}) have polarizations of 0.68, 0.8, 0.81, 0.7,
respectively. Obviously, when the shock nose lies on the
Thomson-scattering surface itself, the polarization of scattered
sunlight is significantly enhanced. In the longitude-elongation
($\varphi-\varepsilon$) maps shown in Figures \ref{Longitude}j-l,
two localized and symmetrical patches of enhanced polarization are
evident. At an elapsed time of 25 hours, one of these patches lies
within the region defined by $40^\circ \le \varepsilon \le
55^\circ$ and $-75^\circ \le \varphi \le -10^\circ$ (Figure
\ref{Longitude}k). Inferring the three-dimensional location of an
interplanetary CME at large elongations using a total intensity
measurement is fraught with uncertainties \cite{Howard-T-A2009a},
which could be significantly reduced through the use of additional
polarization measurements. Locating a CME using measurements of
polarization would require the theoretical calculation of the
Thomson-scattering geometry factors within the CME volume.
However, once the inherent effects of Thomson-scattering are
removed, the total optical intensity can be used to reliably
estimate the mass of the CME. A full sky imager observing
polarized light would not only be capable of monitoring the inner
heliosphere, but would also enable the three dimensional location
of interplanetary disturbances to be established.

The characteristic signature of a large-scale interplanetary
transient in a brightness sky map is a bright arc, followed by a
dark void. The bright arc corresponds to an interplanetary shock.
Such a shock can be further classified either as a blast shock or
a piston-driven shock. A blast shock is generated in response to
an impulsive release of energy in the low corona such as a solar
flare; a piston-driven shock is usually formed ahead of a fast
CME. Coronagraph observations of loop-like transients are
inconsistent with theoretical models of a blast shock wave
\cite{Sime1984}. Here, we consider only the white-light signatures
of the interplanetary shock; the exact nature of the shock is not
of relevance. In practice, features in a brightness sky map need
to be detectable in absolute intensity and exhibit a sharp
gradient along their boundary. Satisfying the criteria of
detectability indicates that an interplanetary disturbance should
be relatively close to the Thomson-scattering sphere, {\it i.e.},
usually within $|\chi^*| \le 30^\circ$ (Figures \ref{Cartoon}c-e).
However, the shock studied in this paper is very strong, described
by initial parameters given in Table \ref{Tab:Shock}. If the shock
was much weaker, the angular extent of its front would be less. In
the case of a weak shock, for an observer at $\varphi=0^\circ$,
the shock front would lie well inside the Thomson-scattering
sphere and would be completely invisible before arriving at 1 AU.
A bright arc observed by an optical imaging device may correspond
to the shock forerunning a CME. During their propagation, CMEs
continuously change their position relative to the
Thomson-scattering surface of a corresponding observer. Therefore,
the white light signature of a travelling CME will evolve
continuously.

\section{Discussions and Summary}\label{Sec:Summary}
In this paper, we present the results of a forward modelling study
of an incident shock in the inner heliosphere. Through combining a
numerical MHD model \cite{Xiong2006a} and Thomson-scattering
theory \cite{Howard-T-A2009a}, we are able to investigate the
causal link between interplanetary dynamics and observable
signatures. By synthesizing white-light observations, we confirm
the significant role played by the Thomson-scattering geometry in
determining the brightness of sky maps at large elongations.
Furthermore, we suggest the use of interplanetary polarization
measurements to help locate interplanetary CMEs; No such
interplanetary observations are currently made. In contrast, the
{\it STEREO}/{\it COR} and {\it LASCO} coronagraphs do have
polarizers, and have been used successfully to investigate CME
orientations, close to the Sun, using a polarimetric
reconstruction technique ({\it e.g.}, \opencite{Moran2010}). As
demonstrated by the proof-of-concept study presented here,
however, interplanetary polarization measurements made possible by
future technological advances would be of significant benefit in
the study of interplanetary transients such as CMEs.

Optical brightness maps contain imprints of various large-scale
interplanetary processes. These processes are usually much more
complex than a simple case of a travelling shock, as studied in
this paper. The use of numerical models is crucial in uncovering
the physics that underlies white-light observations, such that
optical imaging technology can be exploited to its fullest. To our
knowledge, We list some examples below:

\begin{enumerate}
\item Interplanetary coupling of multiple CMEs/shocks.\\
Successive CMEs can interact with each other during their
propagation, and form compound ejecta within 1 AU. A high-speed
CME can drive a fast shock ahead of itself, which is wide in
angular extent than the CME body itself. So CME-shock interaction
is much more likely than CME-CME coupling. Both CME-shock
interaction \cite{Xiong2006a,Xiong2006b} and CME-CME coupling
\cite{Xiong2007,Xiong2009} were theoretically simulated in terms
of interplanetary dynamics and ensuing geoeffectiveness; These
theoretical modelling results are observationally confirmed by a
recent work based on {\it STEREO}/{\it HI} imaging of the
interaction of multiple interplanetary CMEs \cite{Liu2012}. The
most compelling evidence for CME-CME interaction in an optical sky
map is the formation of an intensely bright arc as the CME-driven
shocks completely merge. This bright arc corresponds to the merged
shock front, a region in which the plasma is greatly compressed.
Photospheric data can be used as a driver for numerical MHD models
of the inner heliosphere, to enable realistic simulations of
multiple CMEs to be performed. Following the same method as used
in this paper, optical sky maps can be synthesized and then
compared with observed sky maps. Such a comparison helps validate
the numerical MHD modelling. The use of validated numerical
modelling enables individual CMEs and interaction regions to be
identified within the complex brightness patterns that can be
present in optical sky maps.

\item Interplanetary coupling of a CIR with CMEs/blobs.\\
CIRs are periodic spiral-like structures formed in the inner
heliosphere as a result of compression at the interface between
fast and slow streams. CIRs form where the fast solar wind is
emitted from the rotating Sun, behind the slow solar wind, along
the same solar radial. Like CMEs, although smaller in scale,
plasma blobs contribute to the more transient nature of the inner
heliosphere. Coronagraph observations show that plasma blobs are
intermittently released from the cusps of coronal helmet
streamers, and propagate at the typical slow solar wind speed
\cite{Wang1998}. A CIR can sweep up slow blobs in front, or can
impede the propagation of a following fast CME; Such blobs and/or
CMEs are entrained by the CIR. Plasma blobs that become entrained
within the CIR undergo strong compression such that they are
highly visible in optical sky maps. Blobs entrained at the stream
interface can be identified as recurring patterns of inclined
streaks in time-elongation maps. In the case of CIR-CME
interaction, a CIR could be warped by the entrained CME. CIR-CME
interaction is generally much more disruptive than interaction
between CIRs and blobs. Optical observations of blobs that have
become entrained at the stream interface have been presented by
\opencite{Sheeley2008} and \opencite{Rouillard2008}. CIRs are
inferred to have the spatial morphology of a garden-hose density
spiral, based on their signatures in time-elongation maps.
Interplanetary interaction between CIRs and CMEs/blobs can result
in complex brightness patterns in optical sky maps. The observed
brightness is dictated by multiple factors such as the
Thomson-scattering geometry, the alignment of the spiral CIR along
the LOSs, and local compression within interacting sites.
Numerical models can aid our understanding of these CIR-related
phenomena.

\item Coordinated observations from multiple optical and radio
telescopes.\\
Simultaneous heliospheric imaging from multiple vantage points was
successfully realized with the launch of the twin {\it STEREO}
spacecraft. With stereoscopic imaging, multiple viewing ray-paths
from one observer intersect with those from the other observer.
Thus the three-dimensional distribution of electrons in the inner
heliosphere can be reconstructed using a time-dependent tomography
algorithm (c.f., \opencite{Jackson2003}; \opencite{Bisi2010}). At
times, several IPS ray-paths lie within the imaging FOV such that
CMEs can be simultaneously observed in IPS and white light
\cite{Dorrian2008,Manoharan2010}. The IPS technique can be used to
estimate the location and speed of micro-scale electron density
irregularities \cite{Hewish1964,Coles1989}. If such irregularities
exist within a CME and some assumptions are made about the CME
kinematics, the IPS signals can be used to predict the appearance
of the CME in later optical sky maps \cite{Xiong2011}. It should
be noted that newly constructed low-frequency radio
interferometers such as the LOw Frequency ARray (LOFAR)
\cite{deVos2009} and the Murchison Widefield Array (MWA)
\cite{Lonsdale2009} are proving a major milestone in IPS
technology. Coordinated remote sensing observations in the optical
and radio regimes enable the inner heliosphere to be mapped in
fine detail; Numerical models can guide such endeavors by
suggesting synthesized results beforehand ({\it e.g.},
\opencite{Xiong2011}).
\end{enumerate}

In closing, white-light imaging is a mainstream technology for
remotely sensing the inner heliosphere. However, owing to inherent
geometry effects of Thomson-scattering, deriving the kinematic
properties of interplanetary transients from optical sky maps is
ambiguous at large elongations. The ambiguities arising due to the
Thomson-scattering geometry can be rigorously constrained, if
optical imaging is complemented by other observational techniques,
such as radio imaging and {\it in situ} sampling, and/or by
numerical modelling. More theoretical modelling will be done as a
natural extension to the preliminary results presented in this
paper.

\begin{acks}
This research was supported by a rolling grant from the Science \&
Technology Facilities Council (STFC) to the Aberystwyth
University, UK. Ming Xiong was also partially supported by
Research Fund for Recipient of Excellent Award of the Chinese
Academy of Sciences President's Scholarship.
\end{acks}


\newpage
\section*{}
\begin{table}
\caption{Description of our numerical heliosphere model (Xiong {\it et al.}, 2006a)} 
\label{Tab:model}
\begin{tabular}{cc} \hline %
Mathematical description & MHD equation set \\ \hline %
scheme & Total Variation Diminishing (TVD) + \\  %
& Weighted Essentially Non-Oscillation (WENO) \\ \hline %
type & 2.5 dimensional \\ \hline %
domain & the ecliptic \\ \hline %
coordinate & radial distance $r$, longitude $\varphi$ \\ \hline %
boundary & $25 \le r \le 230$ solar radii, $-180^\circ < \varphi \le 180^\circ$ \\ \hline %
mesh size & $\Delta r = 0.5$ solar radii, $\Delta \varphi = 0.5^\circ$  \\ \hline %
\end{tabular}
\end{table}

\begin{table}
\caption{Prescription of a background solar wind at the inner
boundary of 25 solar radii.}\label{Tab:BG}
\begin{tabular}{cc} \hline %
type & slow solar wind  \\ \hline %
structures & spiral interplanetary magnetic field, \\
 & a heliospheric plasma sheet astride \\ %
 & a heliospheric current sheet (HCS) \\ \hline %
initial longitude of the HCS & $\varphi=\pm \,90^\circ$ \\ \hline %
species & proton $p$, electron $e$ \\ \hline %
number density & $n=n_{\rm p}=n_{\rm e}=550$ cm$^{-3}$ \\ \hline %
radial speed & $v_r=375$ km s$^{-1}$ \\ \hline %
magnetic field strength & $B=400$ nT \\ \hline %
temperature & $T_{\rm p}=T_{\rm e}=9.6 \times 10^5$ K \\ \hline %
plasma beta & $\beta=0.23$ \\ \hline %
alignment condition & $\mathbf{V} \parallel \mathbf{B}$ \\ \hline %
\end{tabular}
\end{table}

\begin{table}
\caption{Initial injection of an incident fast shock via parameter
perturbation at the inner boundary of 25 solar
radii.}\label{Tab:Shock}
\begin{tabular}{cc} \hline %
type & slow solar wind  \\ \hline %
location of shock nose & $\varphi=0^\circ$ \\ \hline %
width of shock front & $12^\circ$ \\ \hline %
temporal duration & 1 hour \\ \hline %
shock speed & 1630 km s$^{-1}$  \\ \hline %
ratio of total pressure & 24  \\ \hline %
\end{tabular}
\end{table}

\begin{table}
\caption{Maximum values of parameters ($z^2 G$, $z^2 G_{\rm R}$,
$z^2 G_{\rm T}$, $n$) along LOSs L1--L7, used for normalization in
Figures \ref{LOS-a} and \ref{LOS-b}. Each LOS is designated a time
$t$, longitude $\varphi$, and elongation $\varepsilon$ in columns
2--4, and is superimposed on Figure \ref{Contour}.}\label{Tab:LOS}
\begin{tabular}{ccccccc} \hline %
LOS & time & longitude & elongation & intensity & Thomson-scattering & electron \\ %
& $t$ & $\varphi$ & $\varepsilon$ & $i$, $i_{\rm R}$, $i_{\rm T}$ & geometry & number \\ %
& (hour) & ($^\circ$) & ($^\circ$) && factors & density $n$ \\ %
&&&&& $z^2 G$, $z^2 G_{\rm R}$, $z^2 G_{\rm T}$ & (cm$^{-3}$) \\ \hline %
L1& 12 & $0$ & $17.5$ & $2.87\times 10^{-27}$ & $2.99 \times 10^{-29}$ & 104.8 \\ \hline %
L2& 25 & $0$ & $40$ & $2.33\times 10^{-28}$ & $6.54 \times 10^{-30}$ & 35.7 \\ \hline %
L5& 37 & $0$ & $90$ & $4.65\times 10^{-29}$ & $2.7 \times 10^{-30}$ & 17.3 \\ \hline %
L6& 37 & $0$ & $155$ & $8\times 10^{-29}$ & $4.92 \times 10^{-30}$ & 17.7 \\ \hline %
L3& 25 & $-45$ & $27$ & $4.96\times 10^{-28}$ & $1.31\times 10^{-29}$ & 41.5 \\ \hline %
L4& 25 & $-45$ & $51$ & $1.46\times 10^{-28}$ & $4.47\times 10^{-30}$ & 32.7 \\ \hline %
L7& 37 & $-90$ & $52$ & $1.11\times 10^{-28}$ & $4.35\times 10^{-30}$ & 25.6 \\ \hline %
\end{tabular}
\end{table}

\newpage
\section*{Figure Captions}
\begin{description}
\item[Figure 1] (a) The Sun as a surface light source, (b)
configuration for the line-of-sight (LOS) integral calculations,
and (c, d, e) typical Thomson-scattering geometries. In panels (a,
c, d, e), points $S$, $Q$, and $O$ indicate the Sun center, a
scattering point, and an observer, respectively. The solar surface
is denoted by a grey semi-circle in panel (a). The $\angle SOQ$ is
defined as elongation $\varepsilon$, $\angle QSO$ as longitude
$\varphi$, and $\angle SQO$ as scattering angle $\chi$. $\chi <
90^\circ$, $\chi = 90^\circ$, and $\chi > 90^\circ$ correspond to
backward, perpendicular, and forward Thomson-scattering,
respectively. The Thomson sphere is the locus of points
corresponding to perpendicular scattering. Point $T$ is the
tangent point on the solar surface from point $Q$, such that
$\angle QTS =90^\circ$. As viewed from point $Q$, the Sun has an
angular half-width of $\angle SQT =\Omega$, and appears as a
luminous disk. As shown in panel (b), for an idealized detector
with a surface area $\delta A$ and a beam size $\delta \omega$,
the recorded brightness in each pixel of the detector is an
integral through a cone determined by the size of the pixel's
point spread function. In panel (b), a local coordinate system is
defined such that its $z$--axis is along the LOS and its
$xy$--plane is in the plane of sky. All diagrams in this figure
are modified from Figures 3, 6, and 8 of
\opencite{Howard-T-A2009a}. The nose of the interplanetary shock
studied in this paper is initially at longitude $\varphi=0^\circ$.
The longitude $\varphi$ of an observer situated to the west (east)
of the shock nose is defined to be positive (negative).
Elongations $\varepsilon$ west and east of the observer's LOS are
positive and negative, respectively.

\item[Figure 2] Thomson-scattering geometry for three observers at
fixed longitudes in the ecliptic. The three observing sites, shown
as white solid tiny circles, are at $\varphi=0^\circ$ (panel (a)),
at $\varphi=-45^\circ$ (panel (b)), and at $\varphi=-90^\circ$
(panel (c)), respectively. In each panel, the ecliptic
cross-section of the corresponding Thomson-scattering sphere is
depicted as a dotted circle. The background image shows the
initial distribution of electron number density $n_0$. Pink and
black solid lines represent the sunward and anti-sunward
interplanetary magnetic field lines, respectively.

\item[Figure 3] Relative electron density enhancement
$(n-n_0)/n_0$ (a, c, e) and radial velocity $v_r$ (b, d, f) for an
interplanetary shock propagating in the ecliptic at three fixed
times, corresponding to 12, 25, and 37 hours. A solid arrow
denotes the LOS from an observer located at a specific longitude
$\varphi$ at a radial distance of 1 AU. Four LOSs (L1, L2, L5, and
L6) are viewed from $\varphi=0^\circ$, two LOS (L3 and L4) from
$\varphi=-45^\circ$, and one LOS (L7) from $\varphi=-90^\circ$.
For each of the three observers, the corresponding
Thomson-scattering sphere is indicated as a dotted circle. Here,
the elongations $\varepsilon$ corresponding to LOSs L1--L7 are
$17.5^\circ$, $40^\circ$, $27^\circ$, $51^\circ$, $90^\circ$,
$155^\circ$, and $52^\circ$, respectively.

\item[Figure 4] Intensities ($i$, $i_{\rm R}$, $i_{\rm T}$),
Thomson-scattering geometry factors ($z^2 G$, $z^2 G_{\rm R}$,
$z^2 G_{\rm T}$), electron number density $n$, polarization $p$,
depth $z$ plotted as a function of modified scattering angle
$\chi^*=90-\chi^\circ$ along L1 (Column (A)), L2 (Column (B)), L5
(Column (C)), and L6 (Column (D)). These are the LOSs that
correspond to an observer at $\varphi=0^\circ$, so the shock is
heading toward the observer. $i$, $i_{\rm R}$, $i_{\rm T}$, $z^2
G$, $z^2 G_{\rm R}$, $z^2 G_{\rm T}$, and $n$ are all normalized
to their respective maximum values along each LOS, as given in
Table \ref{Tab:LOS}. Note that $i=n \,Z^2 G$, $i_{\rm R}=n\, Z^2
G_{\rm R}$, $i_{\rm T}=n\, Z^2 G_{\rm T}$. For $i$ and $n$,
initial and disturbed profiles are depicted as solid and dashed
lines, respectively.

\item[Figure 5] Intensities ($i$, $i_{\rm R}$, $i_{\rm T}$),
Thomson-scattering geometry factors ($z^2 G$, $z^2 G_{\rm R}$,
$z^2 G_{\rm T}$), electron number density $n$, polarization $p$,
depth $z$ plotted as a function of scattering angle
$\chi^*=90-\chi^\circ$ along L3 (Column (A)), L4 (Column (B)), and
L7 (Column (C)). LOSs L3, L4, and L7 correspond to
$\varphi=-45^\circ$, $-45^\circ$, and $-90^\circ$, respectively,
so the shock is propagating off the Sun-observer line. $i$,
$i_{\rm R}$, $i_{\rm T}$, $z^2 G$, $z^2 G_{\rm R}$, $z^2 G_{\rm
T}$, and $n$ are normalized to their respective maximum values, as
given in Table \ref{Tab:LOS}.

\item[Figure 6] The normalization factors $I^*$, $I_{\rm R}^*$,
and $I_{\rm T}^*$ for Figures \ref{Longitude} and \ref{Time}. The
electron number density $n$ is assumed to be dependent on
heliocentric distance $r$ according to the expression $n \propto
r^{-2}$. Assuming that $n \propto r^{-2}$ and using the known
values of the Thomson-scattering geometry factors ($z^2 G$, $z^2
G_{\rm R}$, $z^2 G_{\rm T}$), intensity profiles ($i$, $i_{\rm
R}$, $i_{\rm T}$), calculated along an elongation of
$\varepsilon=30^\circ$, are presented in column (A). Integrated
intensities ($I^*$, $I_{\rm R}^*$, $I_{\rm T}^*$) as a function of
elongation between $0^\circ$ and $180^\circ$ are presented in
column (B). $I^*$, $I_{\rm R}^*$, and $I_{\rm T}^*$ are linearly
scaled such that they are unity at $\varepsilon=6.7^\circ$. In
addition, the polarization distribution $p$ is given in panel (h).

\item[Figure 7] Patterns of brightness $I/I^*$, $I_{\rm R} /I_{\rm
R}^*$, $I_{\rm T} /I_{\rm T}^*$ and polarization $p$, as viewed by
observers at 1 AU and at times of $t=$ 12, 25, and 37 hours, are
presented in the longitude-elongation ($\varphi-\varepsilon$)
parameter space. The dotted lines correspond to $I /I^* = 3.68
\times 10^{-15}$ in panels (a--c), $I_{\rm R} /I_{\rm R}^* =0.68
\times 10^{-15}$ in panels (d--f), and $I_{\rm T} /I_{\rm T}^*
=2.93 \times 10^{-15}$ in panels (g--i). Note that the
normalization factors, $I^*$, $I_{\rm R}^*$, and $I_{\rm T}^*$,
are shown in Figure \ref{calibration}.

\item[Figure 8] Patterns of brightness $I/I^*$, $I_{\rm R} /I_{\rm
R}^*$, $I_{\rm T} /I_{\rm T}^*$, and polarization $p$, as viewed
over the elongation range $\varepsilon=6.7^\circ$ -- $180^\circ$
by observers at 1 AU and at longitudes of $\varphi=0^\circ$,
$-45^\circ$, and $-90^\circ$, are continuously recorded during the
time interval of $t=$ 0 -- 90 hours. The dotted lines correspond
to $I /I^* = 3.68 \times 10^{-15}$ in panels (a--c), $I_{\rm R}
/I_{\rm R}^* =0.68 \times 10^{-15}$ in panels (d--f), and $I_{\rm
T} /I_{\rm T}^* =2.93 \times 10^{-15}$ in panels (g--i). These
dotted lines bound the brightness patterns associated with the
white-light imaging of the travelling shock in the $\varepsilon-t$
parameter space. Note that the normalization factors, $I^*$,
$I_{\rm R}^*$, and $I_{\rm T}^*$, are shown in Figure
\ref{calibration}.
\end{description}

\newpage


\newpage

\newpage

\begin{figure}
\includegraphics[width=0.96\textwidth]{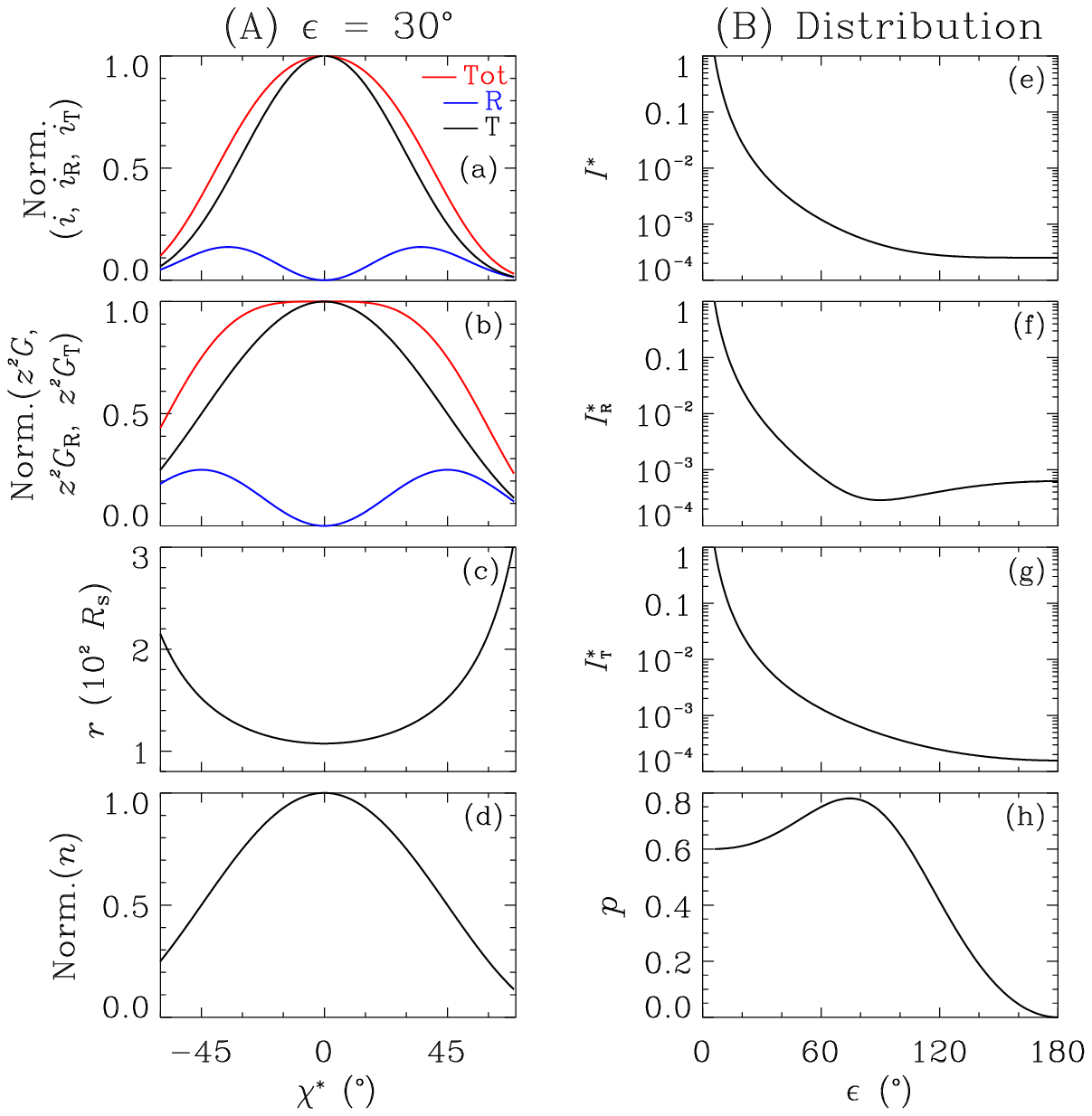}
\caption{} \label{calibration}
\end{figure}

\begin{figure}
\includegraphics[width=0.96\textwidth]{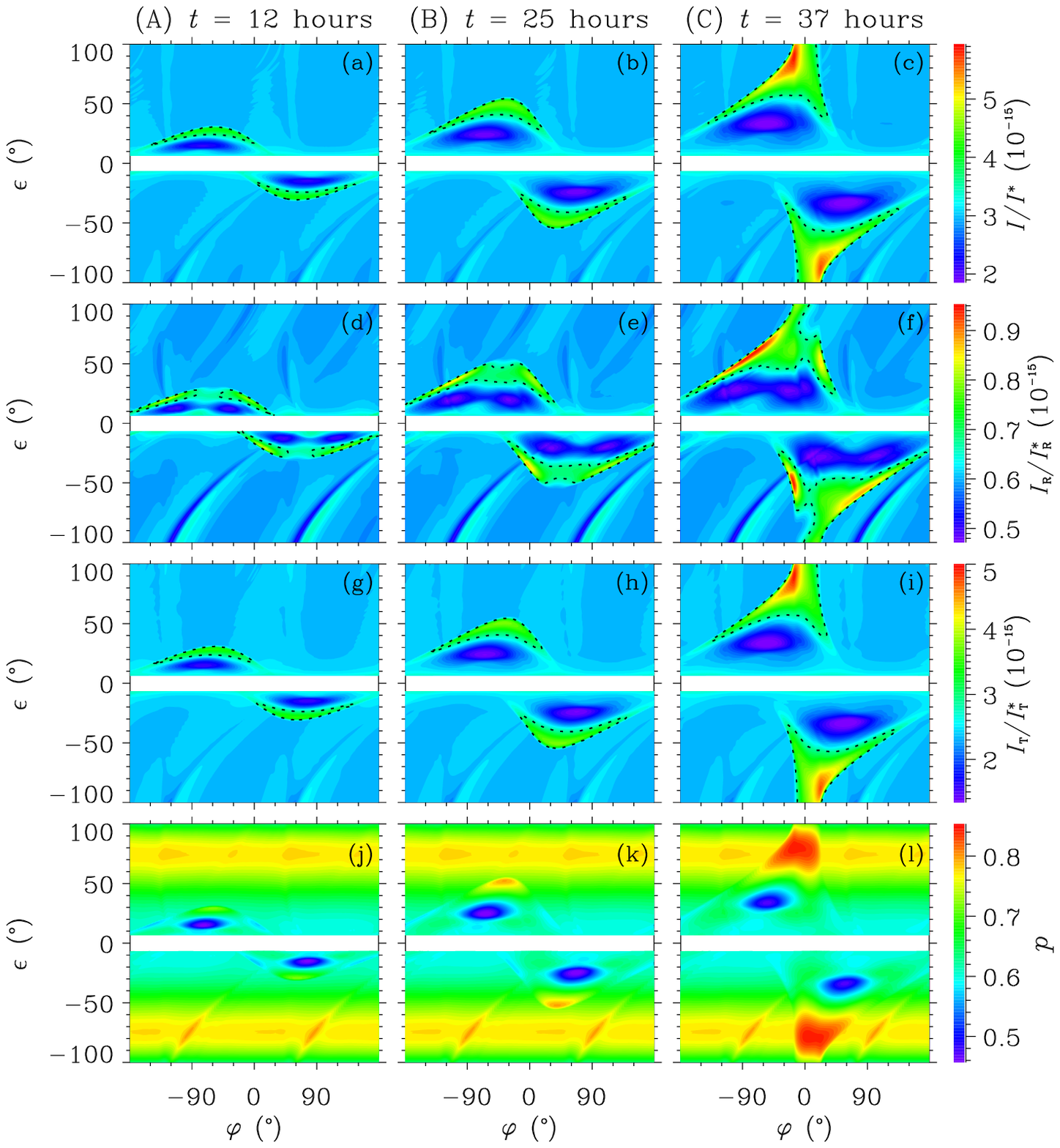}
\caption{} \label{Longitude}
\end{figure}

\begin{figure}
\includegraphics[width=0.96\textwidth]{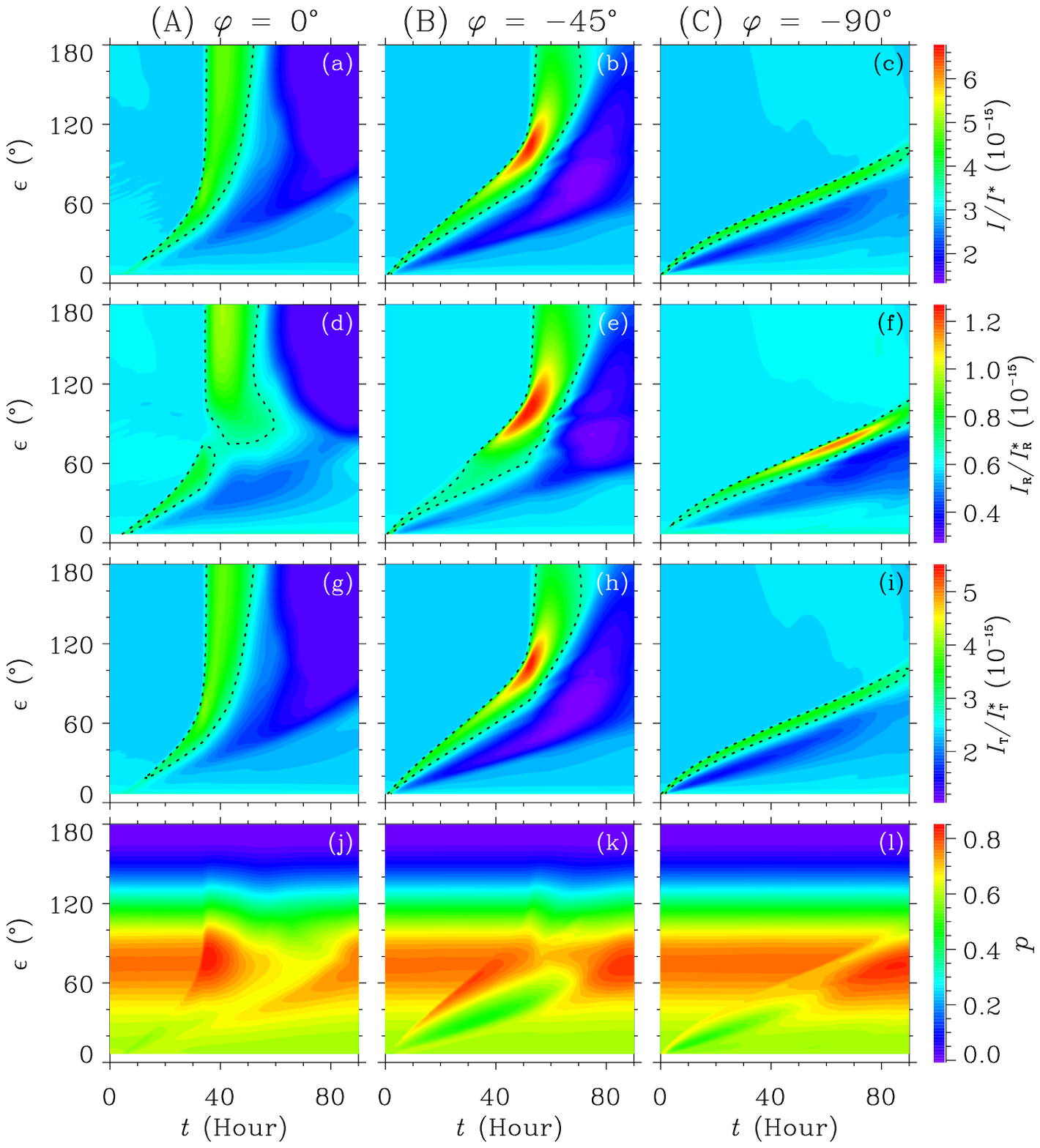}
\caption{} \label{Time}
\end{figure}


\clearpage
\end{article}
\end{document}